\documentclass[sigconf]{acmart}
\usepackage{amsmath}
\usepackage{verbatim}
\usepackage{graphicx}
\usepackage{subfigure}
\usepackage{xcolor}
\usepackage{soul}
\usepackage[utf8]{inputenc}
\usepackage{multirow}
\usepackage{booktabs}

\usepackage{amssymb}
\usepackage[ruled,linesnumbered,vlined]{algorithm2e}
\usepackage{bm}
\usepackage{url}
\usepackage{bbding}
\usepackage[medium,compact]{titlesec}

\usepackage{amsmath,amssymb,amsfonts,bm}

\def\R{\mathrm{I\kern-0.4ex R}}
\def\N{\mathrm{I\kern-0.4ex N}}
\def\E{\mathrm{I\kern-0.4ex E}}
\newcommand{\Set}[1]{\mathcal{#1}}

\AtBeginDocument{%
  \providecommand\BibTeX{{%
    \normalfont B\kern-0.5em{\scshape i\kern-0.25em b}\kern-0.8em\TeX}}}

\begin{document}

\title{Towards Effective Depthwise Convolutions on ARMv8 Architecture}

\author{Ruochen Hao}
\affiliation{%
	\institution{National University of Defense Technology}
	\city{Changsha}
	\country{China}
}
\email{hrc@nudt.edu.cn}

\author{Qinglin Wang}
\affiliation{%
	\institution{National University of Defense Technology}
	\city{Changsha}
	\country{China}
}
\email{wangqinglin_thu@163.com}

\author{Shangfei Yin}
\affiliation{%
	\institution{National University of Defense Technology}
	\city{Changsha}
	\country{China}
}
\email{yin_shangfei@163.com}

\author{Tianyang Zhou}
\affiliation{%
	\institution{National University of Defense Technology}
	\city{Changsha}
	\country{China}
}
\email{ztyemail@126.com}

\author{Siqi Shen}
\affiliation{%
	\institution{Xiamen University}
	\city{Xiamen}
	\country{China}
}
\email{siqishen@xmu.edu.cn}

\author{Songzhu Mei}
\affiliation{%
	\institution{National University of Defense Technology}
	\city{Changsha}
	\country{China}
}
\email{meisongzhu@yeah.net}

\author{Jie Liu}
\affiliation{%
	\institution{National University of Defense Technology}
	\city{Changsha}
	\country{China}
}
\email{liujie@nudt.edu.cn}


\begin{abstract}
	Depthwise convolutions are widely used in lightweight convolutional neural networks (CNNs). The performance of depthwise convolutions is mainly bounded by the memory access rather than the arithmetic operations for classic convolutions so that direct algorithms are often more efficient than indirect ones (matrix multiplication-, Winograd-, and FFT-based convolutions) with additional memory accesses. However, the existing direct implementations of depthwise convolutions on ARMv8 architectures feature a bad trade-off between register-level reuse of different tensors, which usually leads to sub-optimal performance. In this paper, we propose new direct implementations of depthwise convolutions by means of implicit padding, register tiling, etc., which contain forward propagation, backward propagation and weight gradient update procedures. Compared to the existing ones, our new implementations can incur much less communication overhead between registers and cache. Experimental results on two ARMv8 CPUs show that our implementations can averagely deliver 4.88x and 16.4x performance improvement over the existing direct ones in open source libraries and matrix multiplications-based ones in Pytorch, respectively.
\end{abstract}




\maketitle

\section{Introduction}
\label{introduction}
Convolution Neural Networks (CNNs), a class of artificial neural networks, have achieved amazing success in various machine learning tasks, such as image classification \cite{he2015delving}, object detection \cite{ligpu201748}, and medical image diagnostics \cite{singh2020dmenet}. The building blocks of CNNs mainly involve convolutional, pooling, normalization, and fully connected layers. In general, the training and inference of CNNs require a large quantity of computation and memory resource, which are primarily consumed by convolutional layers. The optimization of convolutional layers plays a vital role in improving the performance of CNNs.

Now, many lightweight models have been proposed for mobile computing systems, such as MobileNetV1 \cite{Mobilenetv1}, MobileNetV2 \cite{Mobilenetv2} and MnasNet-A1 \cite{MnasNet-A12019}. These models often consist of a type of convolutions that adopt a single filter for each channel of input feature maps, named depthwise convolutions. In comparison with typical convolutions, depthwise convolutions have much less arithmetic operations and fewer parameters for filters. The sharp reduction of the arithmetic complexity makes the performance of depthwise convolutions is basically bounded by the hierarchical memory bandwidth rather than the peak performance on most platforms \cite{dwaaai2020}.

There are four common methods to perform convolutions, including matrix multiplication \cite{jia2014caffe, ijcnnwang2019}, Winograd \cite{lavin2016fast}, Fast Fourier Transform (FFT) \cite{huangFFT2021} and direct algorithms \cite{zhang2018high}. All the three indirect algorithms introduce the additional transformations, which increase the total overhead of memory access. As a result, the direct algorithm has become a good choice for high-performance depthwise convolutions due to its relatively less memory access.

Although GPUs are the main hardware platforms in deep learning fields, there are many factors to motivate CNNs running on resource-constrained systems including mobile devices (computational and energy constraints) and CPU-based servers (computational constraints relative to popular GPUs) \cite{Survey-2021}. In mobile computing systems, CPUs maybe perform better than GPU in terms of performance and power consumption. Among all the mobile CPUs, the ones based on the ARMv8 architecture have got the largest market share. Moreover, ARMv8 CPUs are rapidly appearing in high performance computing systems, e.g. Mont-Blanc prototype \cite{2016The}, Tianhe-3 prototype \cite{you2019Performance}, and Fugaku supercomputer \cite{fugaku9492415}. Therefore, it's of great significance to optimize direct depthwise convolutions on ARMv8 CPUs.

In deep learning, the two most common data layouts on multi-core CPUs are NHWC (mini-batch, height, weight, channel) and NCHW (mini-batch, channel, height, weight). The latter exhibits better data locality for convolutions so that it is the default layout for Caffe, Mxnet, and Pytorch frameworks \cite{jia2014caffe, mxnet2015, paszke2017pytorch}. But the depthwise convolutions with NCHW layout feature much more irregular  memory access under the vectorized optimization, which largely increase the difficulty of optimization. Existing open-source direct implementations with NCHW layout on ARMv8 architecture are not able to achieve a good balance between the register-level reuse of input and output feature maps tensors shown in Section \ref{background}, and often get sub-optimal performance. Meanwhile, existing researches only involve the forward propagation of depthwise convolutions. This paper focuses on effective direct implementations of depthwise convolutions with NCHW layout, and covers all the procedures for the training and inference.

In order to optimize direct depthwise convolutions, many common techniques like register tiling \cite{registertiling2009}, vectorization, and multi-threading are collaboratively adopted in our work. In comparison with open-source algorithms, the most critical part of our work is how register tiling and implicit padding are applied in the micro-kernel design because it greatly reduces the communication between register and cache under the vectorized optimization with complex access patterns. We believe the application can also inspire similar bandwidth limited algorithms with complex memory access patterns. To the best of our knowledge, this is also the first work which studies direct algorithms for all the three procedures of depthwise convolutions on ARMv8 architecture. The main contributions of this paper can be concluded as follows.


\begin{itemize}
	\item We analyze existing implementations of depthwise convolutions on ARMv8 architecture in detail. It's found that the existing direct implementations focus on the forward propagation, and cannot achieve a good balance among the register-level reuse of all the tensors.

	\item We propose new algorithms with good balanced register-level reuse for depthwise convolutions by means of implicit padding, register tiling, etc., which have less communication overhead between registers and cache, and cover forward propagation, backward propagation and weight gradient update kernels. And, the arithmetic intensity of the new algorithms and the existing ones are introduced to compare their theory performances, with the forward propagation as an example.

	\item All the new algorithms are benchmarked with all different depthwise convolutional layers from MobileNetV1 \cite{Mobilenetv1} and MobileNetV2 \cite{Mobilenetv2} on two ARMv8-based processors: Phytium FT1500A 16-core CPUs and Marvell ThunderX 48-core CPUs. The forward propagation implementations for depthwise convolutions are compared with the direct implementations in open source libraries Tengine, FeatherCNN, ncnn, and ARM Compute Library, while the left two passes are in comparison with the matrix multiplication-based algorithms in the popular deep learning framework Pytorch. The experimental results show that our new algorithms can achieve speedups of up to 2.73x, 6.15x, 7.61x, and 36.38x against Tengine, FeatherCNN, ncnn, and ARM Compute Library respectively, and outperform the matrix multiplication algorithms in all the tests. The optimizations are further confirmed by the training and inference speedup for MobileNetV1 and MobileNetV2, after all the new implementations are used to replace the corresponding kernels in Pytorch.
\end{itemize}

The structure of this paper is as follows. Section ~\ref{background} describes the definition of three procedures for depthwise convolutions, and discusses the relevant existing implementations in detail. Our new implementations of all three procedures are presented in Section ~\ref{implOnArmv8}, and the arithmetic intensities are also analyzed. Section ~\ref{performance} shows the benchmark results on two ARMv8-based CPUs. In Section ~\ref{relatedwork}, we give the prior studies on the optimization of direct convolutions. The conclusion of this paper and the future work can be found in the final section.


\section{Analysis of Existing Implementations}
\label{background}

\subsection{Forward Propagation of Depthwise Convolutions \label{bg.forward}}
The forward propagation of depthwise convolutions takes the input feature maps ($\Set I$) and filters ($\Set F$), and produces the output feature maps ($\Set O$). In NCHW layout, these tensors are expressed as $\Set I[N][C][H_i][W_i]$, $\Set F[C][H_f][W_f]$ and $\Set O[N][C][H_o][W_o]$. Thus, depthwise convolution is defined by
\begin{equation}\label{eq.conv1}
	\begin{aligned}
		\Set O_{n, c, h_o, w_o} =   \sum_{h_f=0}^{H_f-1}{\sum_{w_f=0}^{W_f-1}{ }}
		(\Set I_{n, c, h_o \times s + h_f - p_t, w_o \times s + w_f - p_l} \\ \times  \Set F_{c, h_f, w_f}),
	\end{aligned}
\end{equation}
where $0 \le n < N$, $0 \le c < C$, $0 \le h_o < H_o$, $0 \le w_o < W_o$, $N$ is the mini-batch size, $C$ is the number of channels, $H_{i/o/f}$ and $W_{i/o/f}$ denote the spatial height and width, $p_{t/l}$ refers to the padding size in the spatial dimension, and  $s$ is the stride size. In this paper, we mainly focus on depthwise convolutions with  $H_f \times W_f = 3 \times 3$ and  $s \in \{ 1, 2\} $, which are the most common cases in lightweight models.
From the equation \ref{eq.conv1}, it can be found that the forward propagation is actually performing $N$ batched matrix-vetor multiplications. During the computation, the filters are the shared tensors and often small enough to be kept in the on-chip memory all the time. The input feature maps are streamed into the on-chip memory, and then the produced output feature maps are streamed back into the main memory. In other words, there is little space to optimize the access to the main memory, and we will mainly study the communication between cache and register in the following.
\label{tengine.analysis}
\begin{figure}[H]
	\setlength{\abovecaptionskip}{-0.2cm} 
	\setlength{\abovecaptionskip}{0.0cm} 
	\centering
	\includegraphics[width=1.00\linewidth]{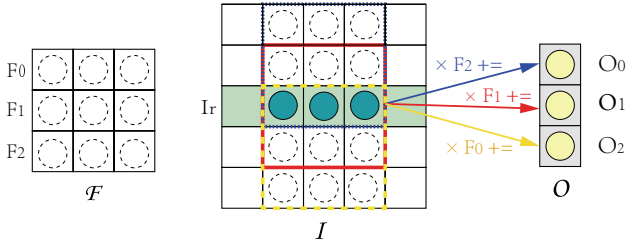} 
	\caption{Data reuse in Tengine where $H_f \times W_f = 3 \times 3$ and the stride size is 1.}
	\label{fig.tengine}
\end{figure}
We tested existing implementations for forward propagation of depthwise convolution in Tengine, FeatherCNN, ncnn and ARM Compute Library on two ARMv8 CPUs, and found that the best performance was achieved in most cases by Tengine.
Tengine's implementation is illustrated in Fig. \ref{fig.tengine}.
In ARMv8 architecture, when the width of vector units is 128-bit, each vector unit can deal with 4 single precision numbers simultaneously. In Tengine, each row of $\Set I$ is only loaded one time from cache to vector registers while each row of $\Set O$ is repeatedly loaded from cache to vector registers. For each sample, the elements of $\Set F$ are also loaded into vector registers one time. In other words, in height direction, the elements of $I$ in registers are reused $H_f$ times while there is no reuse for the ones of $\Set O$. For example, $I_r$ are reused three times to get $O_{0-2}$ after loaded into registers, shown in Fig. \ref{fig.tengine}. The elements of $\Set O_{0-2}$ are loaded into vector registers twice and stored back into cache three times. Therefore, we can find the philosophy in Tengine is reusing the elements of $\Set I$ in registers as much as possible so that the communication overhead of $\Set O$ between registers and cache are largely increased. The total communication between registers and cache is about
$TC_{tg}$ = $(N \times C \times H_i \times W_i + N \times C \times H_f \times W_f + 5 \times N \times C \times H_o \times W_o) \times 4$ Bytes.
\subsection{Backward Propagation of Depthwise Convolutions \label{bg.backward}}
The backward propagation of depthwise convolutions takes the output gradient tensor ($d\Set O$) and the filters ($\Set F$), and produces the input gradient tensor ($d\Set I$). The computation process can be described by the following equation:
\begin{eqnarray}\label{eq.iconv}
	\begin{aligned}
		d\Set I_{n, c, h_o \times s + h_f-p_t, w_o \times s + w_f-p_l} \mathrel{+}= d\Set O_{n, c, h_o, w_o} \\ \times  \Set F_{c, h_f, w_f},
	\end{aligned}
\end{eqnarray}
where $0 \le n < N$, $0 \le c < C$, $0 \le h_o < H_o$, $0 \le w_o < W_o$.

From the equation \ref{eq.iconv} we can see that the backward propagation is similar to the forward propagation, so the optimizing strategy is still to reduce the communication between cache and register. Most deep learning libraries and frameworks on ARMv8 architectures usually implement the backward propagation and weight gradient update of depthwise convolutions through the matrix multiplications based algorithms, which complete the passes by executing a batch of matrix multiplications. In the backward propagation, the matrix multiplication routine is carried out on the $C \times M_m \times K_m $ matrix $\Set {F'}$ and the $C \times N_m \times K_m $ matrix $d\Set {O'}$ to get the $C \times M_m \times N_m $ matrix $d\Set {I'}$, where $M_m = H_f \times W_f$, $N_m = N \times H_o \times  W_o$ and $K_m = 1$. Finally, $d \Set I$ can be obtained by performing the column-to-image transformation on matrix $d\Set {I'}$, which incurs huge communication overhead between different levels in the memory hierarchy. Therefore, the matrix multiplications based backward propagation often can't achieve the optimal performance.



\subsection{Weight Gradient Update of Depthwise Convolutions \label{bg.weight}}
The weight gradient update of depthwise convolutions takes the input tensor ($\Set I$) and the output gradient tensor ($d\Set O$), and produces the weight gradient tensor ($d\Set F$). The computation process can be described by the following equation:
\begin{equation}\label{eq.kconv}
	\begin{aligned}
		d\Set F_{c, h_f, w_f} = \sum_{n=0}^{N-1}{\sum_{h_o=0}^{H_o-1}{\sum_{w_o=0}^{W_o-1}{}}}
		(\Set I_{n, c, h_o \times s + h_f - p_t, w_o \times s + w_f - p_l} \\ \times  d\Set O_{n, c, h_o, w_o}),
	\end{aligned}
\end{equation}
where $0 \le n < N$, $0 \le c < C$, $0 \le h_o < H_o$, $0 \le w_o < W_o$.

In the matrix multiplication based algorithm of weight gradient update, the output gradient tensor is reshaped into $C \times N_m \times K_m$ matrix$d\Set O'$, the input tensor ($\Set I$) is lowered into a $C \times M_m \times N_m$ Toeplitz matrix $\Set I'$ by image-to-column operations,  and the generated $C \times M_m \times K_m$ matrix is finally reshaped back into $d\Set F$, where $M_m = H_f \times W_f$, $N_m = N \times H_o \times  W_o$ and $K_m = 1$. Therefore, the algorithm also brings about the huge redundant memory access for the weight gradient update, which incurs performance penalties.

\section{Our Approach}
\label{implOnArmv8}
In this section, we will illustrate how the optimizing techniques are applied to the forward propagation, backward propagation and weight gradient update procedures of direct depthwise convolutions, and analyze the arithmetic intensity.

\subsection{Forward Propagation Implementation}
In depthwise convolutions, the elements of both $\Set I$ and $\Set O$ can be reused up to $H_f \times W_f$ times in registers.
Each core in ARMv8 CPUs has only 32 vector registers so that we may need to load the same elements multiple times from cache
to registers. As analyzed in Section \ref{tengine.analysis}, Tengine loads the elements of $\Set O$ multiple times while
maximizing the reuse of the elements of $\Set I$. However, the repeated loading of $\Set O$ is always accompanied by the
repeated reading and storing of $\Set O$ from registers to cache due to the accumulation.
Therefore, our approach chooses to maximize the reuse of $\Set O$ in registers, and uses implicit padding, vectorization and
register tiling techniques to maximize the reuse of $\Set I$ in registers, so that the total communication between
registers and cache can be largely reduced. Our direct implementation for the forward propagation is shown in Algorithm \ref{forward.our}.

\subsubsection{Implicit Padding}
\begin{figure}[!htbp]
	\setlength{\abovecaptionskip}{-0.2cm} 
	\setlength{\abovecaptionskip}{0.0cm} 
	\centering
	\includegraphics[width=0.9\linewidth]{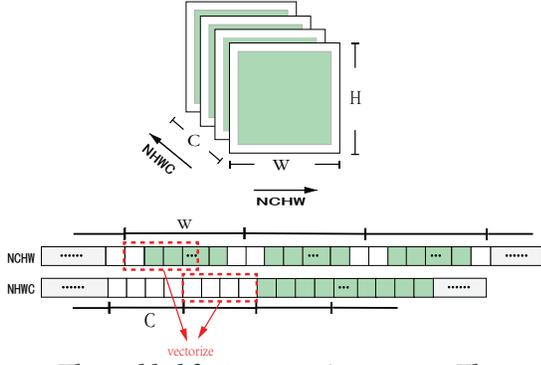} 
	\caption{The padded feature map in memory. The grey and green blocks represent padded elements and input elements respectively.}
	\label{layout}
\end{figure}

Compared to depthwise convolutions with NHWC layout, the ones with NCHW layout features irregular memory access, especially when dealing with padding. Fig .\ref{layout} shows the way how the padding is carried out into the input feature maps with NCHW and NHWC layout. The vectorization is performed along width dimension and channel dimension in NCHW and NHWC data layout, respectively. The red blocks in Fig .\ref{layout} indicate that a vector may consist of input and padding elements simultaneously in NCHW layout, while a vector includes only padding elements or input elements from VL(vector length) channels in NHWC layout. Therefore, it's more difficult and expensive to deal with the padding under the vectorization optimization with NCHW layout. There are two common methods for padding. One method is explicitly padding input feature maps into a temporary space before computation, adopted by ncnn\cite{ncnn2021} and FeatherCNN\cite{feathernn2021}. The other is that the padding is implicitly done through data movement between registers during computation.
In comparison with the former method, the latter brings the overhead of data movement in registers.
However, the latter has only a half of the communication overhead between cache and registers in the former when only padding is considering.
As the performance of depthwise convolution is mainly limited by memory access latency, our approach adopts the implicit padding to minimize the overhead of cache access in padding, shown in lines \ref{padding_start} - \ref{padding_end} of Algorithm \ref{forward.our}.


\begin{algorithm}[htb!]
	\SetAlgoNoLine
	\SetAlgoNoEnd
	\DontPrintSemicolon
	\SetKwFunction{Union}{Union}
	\SetKwFunction{FindCompress}{FindCompress}
	\SetKwFunction{kernel}{KernelHrxWr}
	\SetKwProg{Fn}{Function}{:}{}
	\SetKwInOut{Input}{input}\SetKwInOut{Output}{output}
	\Input{Input feature maps $\Set I$, Filter $\Set F$}
	\Output{Output feature maps $\Set O$}

	\ForPar{$n= 0 \colon 1 \colon N$\label{thread1}} {
		\ForPar{$co = 0 \colon C_{b} \colon C $\label{thread2}}  {
			\For{$c = co \colon 1 \colon co + C_{b} $}  {
				$vf_{0:H_f-1} = simd\_load(\Set F_{c,0:H_f-1,0})$\; \label{load_kernel}
				\For{$h_o = 0 \colon H_{r} \colon H_{o}$ \label{thread_ho}} {
					\For{$w_o = 0 \colon W_{r} \colon W_o$}  {
						\kernel{n,c,$h_o$,$w_o$}
					}
				}
			}
		}
	}
	\Fn{\kernel{n,c,$h_o$,$w_o$}\label{twok_start}}{
	$W_n = {W_{r}}/{VL}$\;
	$h_{i} = h_o \times s - p_t$\;
	$w_{i} = w_o \times s - p_l$\;
	\For{$r = 0 \colon 1 \colon H_f + (H_{r} - 1) \times s$ \label{loadi}} {
	Step1: load $W_f + (W_r-1) \times s$ elements from $\Set I_{n,c,h_{i} + r, w_i}$\label{padding_start}\;
	Step2: extract padding/loaded elements into $vi_{0:W_n \times W_f-1}$\label{padding_end}\;
	\For{$h = 0 \colon 1 \colon H_{r}$} {
		\For{$h_f = 0 \colon 1 \colon H_f$}{
			\lIf{$h \times s + h_f \mathrel{!}= r$}{
				break
			}
			\For{$w = 0 \colon 1 \colon W_n$}  {
				\For{$q = 0 \colon 1 \colon W_f$}  {
					$vo_{h \times W_n + w} = simd\_fma(vo_{h \times W_n + w}, vi_{w \times W_f + q}, vf_{h_f}[q])$
				}
			}
		}
	}
	}
	\For{$h = 0 \colon 1 \colon H_{r}$} {
		Store $vo_{h \times W_n:(h+1) \times W_n-1}$ back into $\Set O_{n,c,h_o + h,wo} \label{store_output}$
	}\label{twok_end}
	}
	\caption{Our Direct Implementation for forward propagation of Depthwise Convolutions}
	\label{forward.our}
\end{algorithm}
\subsubsection{Vectorization and Register Tiling}
We further utilize vectorization and register tiling techniques to increase register-level data reuse.
In convolutional operations, the filter firstly slides along the width dimension, so that the adjacent convolutions in the same row involve two overlapped regions from the input feature. To reduce the redundant loads, we vectorize the convolution operations in width direction, which divides the convolution operations in a row into groups of $VL$(vector length).
Next, we divide the elements of $\Set O$ into blocks of $H_r \times W_r$ size in the height and width dimensions to fix the data used in the computation of the basic block in register.
As the vectorization is carried out in the width dimension, $W_r$ must be a multiple of $VL$. At the same time, $H_r$ and $W_r$ are also limited by the total number of vector registers in ARMv8 CPUs. The kernel function with $H_r \times W_r$ tiling is shown in lines \ref{twok_start} - \ref{twok_end} of Algorithm \ref{forward.our}. The elements of $\Set F$ are loaded into registers $vf_{0:H_f-1}$ in advance in lines \ref{load_kernel}. There are $H_r \times W_r/VL$ registers for the $H_r \times W_r$ block of $\Set O$, namely, $vo_{0:H_r \times W_n-1}$. And $vo_{0:H_r \times W_n-1}$ are reused $H_f \times W_f$ times in the kernel function, and are only stored back into cache once. $H_f + (H_{r} - 1) \times s$ rows of $\Set I$ are involved to compute the  $H_r \times W_r$ block of $\Set O$, and each row will be extracted into $W_f \times W_r/VL$ vectors, namely $vi_{0:W_f \times W_r/VL - 1}$, through register manipulations, and the padding operation is performed implicitly in this step.

\begin{figure}[!htbp]
	\setlength{\abovecaptionskip}{-0.2cm} 
	\setlength{\abovecaptionskip}{0.0cm} 
	\centering
	\includegraphics[width=0.9\linewidth]{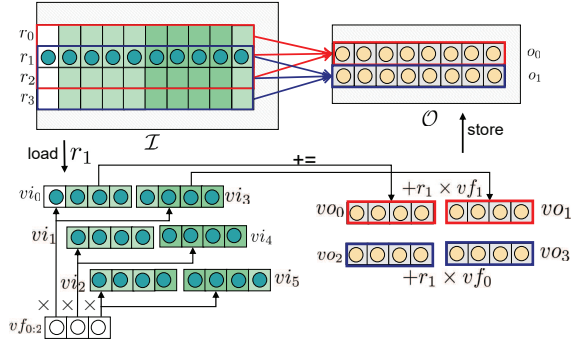} 
	\caption{A working example of our forward propagation implementation of stride-1 depthwise convolutions  on ARMv8 CPUs, where $VL=4$, $H_f \times W_f = 3 \times 3$ and $H_r \times W_r = 2 \times 8$.}
	\label{vectorize1}
\end{figure}

\begin{figure}[!htbp]
	\setlength{\abovecaptionskip}{-0.2cm} 
	\setlength{\abovecaptionskip}{0.0cm} 
	\centering
	\includegraphics[width=0.90\linewidth]{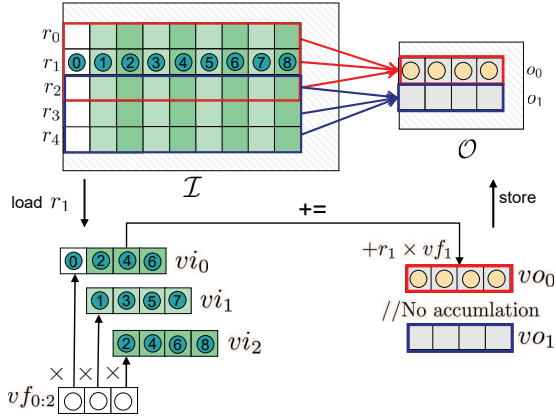} 
	\caption{A working example of our forward propagation implementation of stride-2 depthwise convolutions on ARMv8 CPUs, where $H_f \times W_f = 3 \times 3$ and $H_r \times W_r = 2 \times 4$.}
	\label{vectorize2}
\end{figure}

For a more intuitive description of the computation procedure, we will go through the examples depicted in Fig. \ref{vectorize1} and Fig. \ref{vectorize2}. Fig. \ref{vectorize1} illustrates the case of unity stride. Without loss of generality, $H_r$ and $W_r$ are set to $2$ and $8$ in Fig. \ref{vectorize1}. We first load a row ($r_1$) of input and extract it into $vi_{0:5}$ along with padding elements. Then we multiply the corresponding elements of $vf_0$ and $vf_1$ to $vi_{0:5}$ as indicated by the red and blue arrows, and the generated results are accumulated to the output vectors $vo_{0:1}$ and $v0_{2:3}$. The elements in $vo_{0:3}$ are stored back until the final results of depthwise convolutions are acquired. Thus, when the stride is 1, the vectorization and register tiling strategy allow the elements of $r_1$ to be reused almost three times in width dimension and to be reused twice in height dimension, and the elements in $vo_{0:3}$ are reused 9 times. The case of stride $2$ is shown in Fig. \ref{vectorize2}. The elements of the loaded row ($r_1$) are extracted into $vi_{0:2}$ with a stride of 2, as indicated by the different colors and numbers. The multiplication and accumulation operations follow the same process as the former case. The strategies play the same role in the case of stride 2, but exhibit much less reuse times on account of larger stride size. For example, in a kernel's computation, only the row $r_2$ is reused twice and the other rows $r_{0:1}$ and $r_{3:4}$ are used only once. In total, the register-level data locality is determined by the tiling size $H_r \times W_r$ and the stride size.

The tiling size is mainly determined by maximizing attainable data locality in registers, and also limited by the total number of registers. As far as stride $1$ is concerned, we adopt $4 \times 4$ tiling size in most cases. As the boundary part often requires additional logical judgement and can not be efficiently vectorized, the overhead of the boundary part increases when the size ($H_i \times W_i$) of feature maps become small. When the height of input size decreases to some threshold value, the implementation with $4 \times 4$ tiling size can only achieve suboptimal performance. Therefore, we lower the tiling size in height to $2$ and increase the tiling size in width to $8$ simultaneously to invoke the $2\times 8$ kernel to handle the boundary. When it comes to stride-$2$ kernels, we get the best performance with $1 \times 4$ tiling size.

Additionally, register tiling also can increase the number of operations which can be processed in parallel and filled into the pipeline of ARMv8 CPUs, so that the latency of instructions can be efficiently hidden. It is worth noting that the function \emph{Kernel$H_r \times W_r$} is implemented in assembly language, and all loops are unrolled.

\subsubsection{Parallelism}
Apart from improving micro-kernel performance through aforementioned techniques, our approach can also increase the thread-level parallelism.
From Algorithm \ref{forward.our} we can see that the outer-most two loops(Lines \ref{thread1} - \ref{thread2}) are parallelized by default,
which provides $N \times C$ work items. The blocking parameter $C_b$ can be adjusted according to the size of input tensor and thread number.
However, the default parallelism strategy is not sufficient when the number of threads is greater than $N \times C$.
Since the computing of output($\Set O$) blocks are independent, we further parallelize loop $h_o$(Line \ref{thread_ho}) to fully utilize processor's parallel processing capacity.
\subsection{Backward Propagation Implementation}
\begin{figure}[!htbp]
	\setlength{\abovecaptionskip}{-0.2cm} 
	\setlength{\abovecaptionskip}{0.0cm} 
	\centering
	\includegraphics[width=0.99\linewidth]{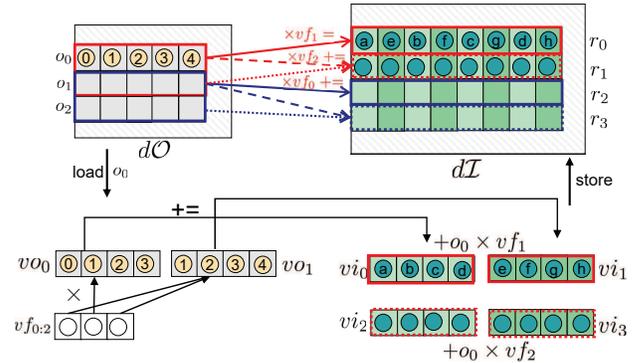}
	\caption{A working example of our backward propagation implementation of stride-2 depthwise convolutions on ARMv8 CPUs, where $H_f \times W_f = 3 \times 3$ and $H_r \times W_r = 2 \times 8$}
	\label{back_fig}
\end{figure}
In this part, we will introduce our implementation of backward propagation.
Backward propagation is described by equation \ref{eq.iconv} mentioned in section \ref{bg.backward}.
By setting $h_i=h_o+h_f-p_t$, $w_i=w_o+w_f-p_l$, we can rewrite the equation
as $dI[n][c][h_i][w_i] \mathrel{+}= dO[n][c][(h_i-h_f+p_t)/s][(w_i-w_f+p_l)/s] \times F[c][h_f][w_f]$.
In the case of unity stride, the equation is equivalent to
$dI[n][c][h_i][w_i] \mathrel{+}= dO[n][c][h_i-h_f+p_t][w_i-w_f+p_l] \times F[c][h_f][w_f]$.
By setting $h_f'=-h_f$, $w_f'=-w_f$, $p_t'=-p_t$, $p_l'=-p_l$, the equation can be rewrite as
$dI[n][c][h_i][w_i] \mathrel{+}= dO[n][c][h_i+h_f'-p_t'][w_i+w_f'-p_l'] \times F'[c][h_f'][w_f']$,
where $F'[c][-h_f][-w_f] = F[c][h_f][w_f]$, which is the same as the equation of forward propagation.
So we can take $d\Set O$ and $\Set F'$ as inputs and invoke the kernel presented
in Algorithm \ref{forward.our} to implement efficient backward propagation.
It is worth mentioning that $\Set F'$ is the rotation of $\Set F$ by 180 degrees.

However, the problem is more complicated in the second case and we will illustrate it's micro-kernel design in detail.
When s=2, we have $h_o=(h_i+p_t-h_f)/2$ and $w_o=(w_i+p_l-w_f)/2$.
In order for the value of $h_i+p_t-h_f$ to be divisible by 2, the parity of $h_i$ and $p_t-h_f$ must be the same.
That is, when $h_i$ is even, the value of $p_t-h_f$ must also be even;
similarly, when $w_i$ is odd, the value of $p_t-h_f$ must also be odd.
Since $p_t$ and $p_l$ are constants, the elements of $\Set F$ involved in the computation of $d\Set I_{n,c,h_i,w_i}$
are decided by the parity of $h_i$ and $w_i$.
Therefore, there are 4 cases of the computation formula for $d\Set I$, which depends on the parity of $hi$ and $wi$.
Besides, every other elements($d\Set I$) in a row have the same parity, thus they share the same formula and can be vectorized.
Through register tiling strategy, we divide elements of $d\Set I$ into blocks of $H_r \times W_r$ size in the height and width dimensions.
As the computations of every other elements ($d\Set I$) are vectorized in width dimension, $W_r$ must be a multiple of $2 \times VL$.
Fig. \ref{back_fig} shows an example of our approach with $2 \times 8$ tiling size, where $p_t$ and $p_l$ are set to $1$.
For brevity, we omit $h_i$, and derive the following equation of $d\Set I$:
\begin{small}
	\begin{equation}
		\hspace{-2mm}
		d\Set I_{n,c,*,w_i}=\left\{
		\begin{array}{rc}
			\hspace{-2mm} d\Set O_{n,c,*,w_i/2} \times \Set F_{c,*,1}                                                                             & \hspace{-2mm} w_i=2k   \\
			\hspace{-2mm} d\Set O_{n,c,*,\lfloor w_i/2\rfloor } \times \Set F_{c,*,2} + d\Set O_{n,c,*,\lceil w_i/2\rceil } \times \Set F_{c,*,0} & \hspace{-2mm} w_i=2k+1 \\
		\end{array} \right.
		\label{inferback}
	\end{equation}
\end{small}
From the equation \ref{inferback}, we can deduce that every x elements in a row of $d\Set I$
are related to x/2+1 elements in a row of $d\Set O$.
The equation \ref{inferback} also applies for height dimension.The relations between rows
of $d\Set O$ and $d\Set I$ are indicated by the arrows in the top of Fig. \ref{back_fig}.
The detailed computation procedures are described in the following.
The elements of $\Set F$ are loaded into registers $vf_{0:2}$ in advance.
We load a row($o_0$) of $d\Set O$ and extract it into $vo_{0:1}$ by implicitly padding and register manipulations.
As illustrated by the arrows from $vf_{0:2}$, $vo_{0:1}$ are then multiplied by the corresponding elements of
$vf_{1}$ and $vf_{2}$, and the results are accumulated into $vi_{0:1}$ and $vi_{2:3}$ respectively.
The temporary results in $vi_{0:3}$ are kept in the registers until the final results are obtained.
Since the computation of $d\Set I_0$ is finished in Fig. \ref{back_fig}, the $vi_{0:1}$
are written back to memory with a stride of 2, which can by indicated by the letters labeled on the elements.
We tried some common tiling sizes and adopted the $6 \times 8$ for its higher performances in most layers.

\subsection{Weight Gradient Update Implementation}
\begin{figure}[!htbp]
	\setlength{\abovecaptionskip}{-0.2cm} 
	\setlength{\abovecaptionskip}{0.0cm} 
	\centering
	\includegraphics[width=0.99\linewidth]{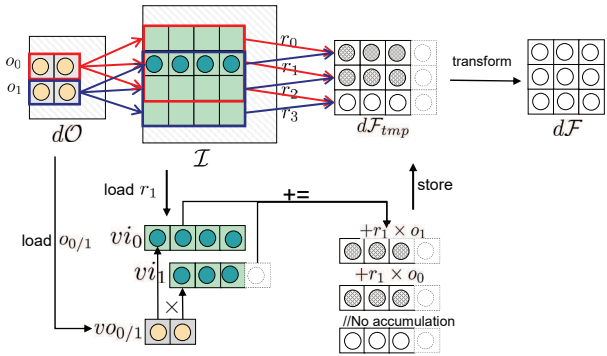} 
	\caption{A working example of our weight gradient update propagation implementation of stride-1 depthwise convolutions on ARMv8 CPUs, where $H_f \times W_f = 3 \times 3$, and $H_r \times W_r = 2 \times 2$}
	\label{fig.weight1}
\end{figure}
\begin{figure}[!htbp]
	\setlength{\abovecaptionskip}{-0.2cm} 
	\setlength{\abovecaptionskip}{0.0cm} 
	\centering
	\includegraphics[width=0.99\linewidth]{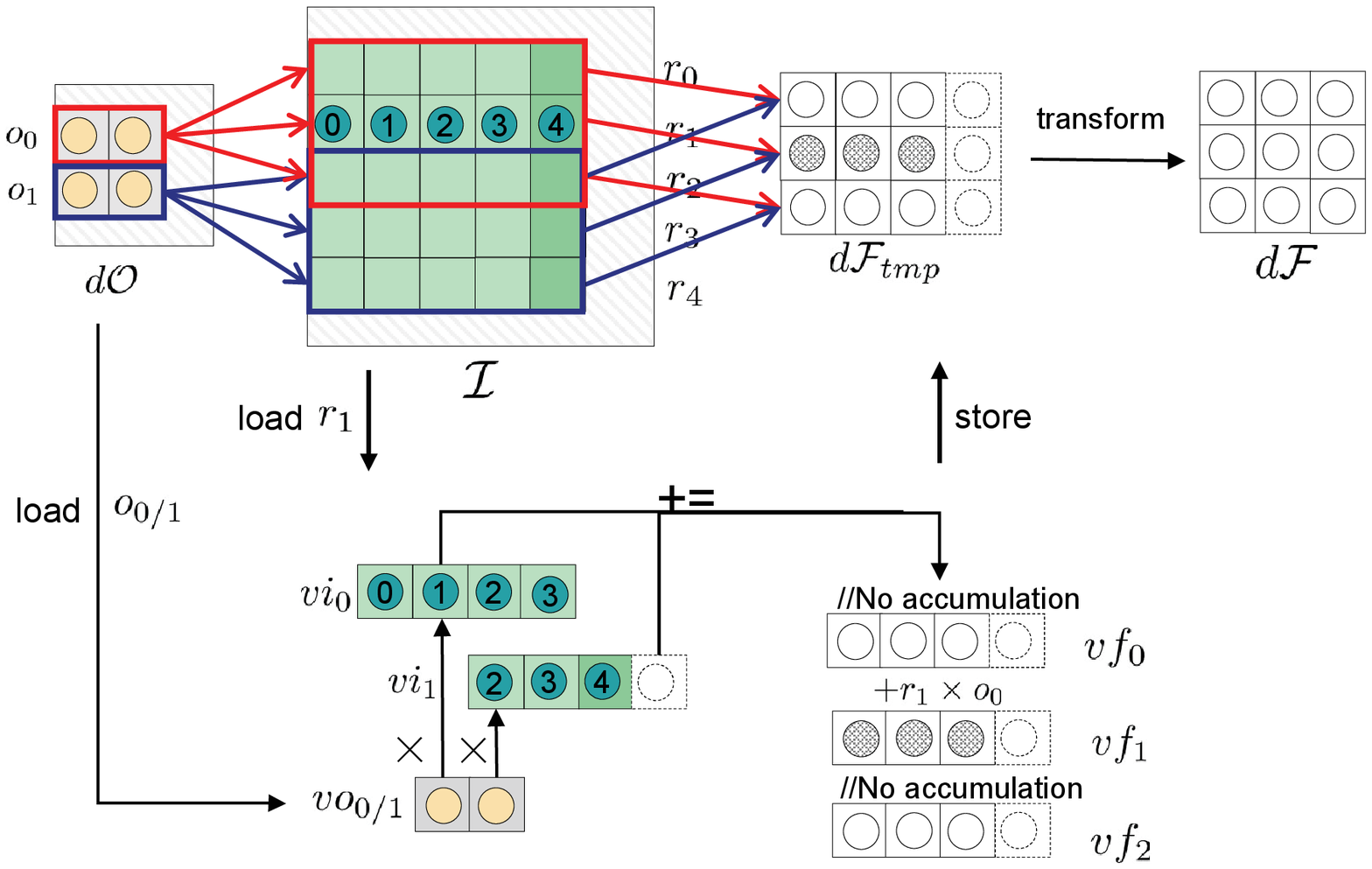} 
	\caption{A working example of our weight gradient update propagation implementation of stride-2 depthwise convolutions on ARMv8 CPUs, where $H_f \times W_f = 3 \times 3$, and $H_r \times W_r = 2 \times 2$}
	\label{fig.weight2}
\end{figure}
\begin{algorithm}[htb!]
	\SetAlgoNoLine
	\SetAlgoNoEnd
	\DontPrintSemicolon
	\SetKwData{Left}{left}\SetKwData{This}{this}\SetKwData{Up}{up}
	\SetKwFunction{Union}{Union}
	\SetKwFunction{FindCompress}{FindCompress}
	\SetKwFunction{wkernel}{Kernel}
	\SetKwProg{Fn}{Function}{:}{}
	\SetKwInOut{Input}{input}\SetKwInOut{Output}{output}
	\Input{Input feature maps $\Set I$, Gradient Output $d\Set O$}
	\Output{Weight Gradient $d\Set F$}

	alloc $\Set F\_tmp[C][H_f][VL]$\;
	\ForPar{$c = 0 \colon 1 \colon C $ \label{exchangec}}  {
		\For{$n= 0 \colon 1 \colon N$ \label{exchangeb}} {
			\For{$h_o = 0 \colon H_r \colon H_o$ \label{block0}} {
				\For{$w_o = 0 \colon W_r \colon W_o$ }  {
					\wkernel{n,c,$h_o$,$w_o$} \label{block1}
				}
			}
		}
		Store $vf_{0:H_f-1}$ back to $\Set F\_tmp_{c,0:H_f-1,0}$\\
	}
	Store $\Set F\_tmp$ back to $d\Set F$\\

	\Fn{\wkernel{n,c,$h_o$,$w_o$}}{\label{kernelstart}
		$W_n$ = $W_r/VL$\\
		$h_i$ = $h_o \times s - pt$\\
		$w_i$ = $w_o \times s - pl$\\
		\For{$h = 0 \colon 1 \colon H_r$ \label{loadblock0}} {
			$simd\_load$ $W_r$ elements from $d\Set O_{h_o+h,w_o}$ to $vo_{h \times W_n:(h+1) \times W_n-1}$\\ \label{loadblock1}
		}
		\For{$r = 0 \colon 1 \colon H_r \times s + 1$} {
			Step1:load $(W_r-1) \times s + W_f$ elements from $\Set I_{n,c,h_i+r,w_i}$\\
			Step2:extract loaded elements($\Set I$)/padding elements into $vi_{0:W_r-1}$\\
			\For{$h = 0 \colon 1 \colon H_r$} {
				\For{$h_f = 0 \colon 1 \colon H_f$}{
					\If{$h \times s + h_f != r$} {break}
					\For{$w = 0 \colon 1 \colon W_r$} {
						$wn = w / VL$\\
						$q = w \% VL$\\
						$vf_{h_f}$ = $simd\_fma(vf_{h_f}, vi_{w}, vo_{h\times W_n + wn}[q]])$\\ \label{fma}
					}
				}
			}
		}
	}
	\caption{Our Implementation of Weight Gradient Update}
	\label{our.weight}
\end{algorithm}
In this sub-section, we will present our implementation of weight gradient update.
Our implementation is described in Algorithm \ref{our.weight}.
As shown in equation \ref{eq.kconv}, the gradient weight feature maps in the
same channel $c$ will be added up to  $d\Set F_{c,0,0}$, so the loop c is parallelizable(Line \ref{exchangec}).
The elements of $d\Set O$ are divided into blocks of $H_r \times W_r$ size
in the height and width dimensions(Lines \ref{block0}-\ref{block1}) and
each block is processed by function \emph{Kernel}(Lines \ref{kernelstart}-\ref{fma}).
At the beginning of \emph{Kernel}, elements($d\Set O$) in the block are loaded into registers(Lines \ref{loadblock0}-\ref{loadblock1})
and are kept until all the related elements of $\Set I$ are multiplied by it.
When the filter size is $H_f \times W_f$,
in order to process vectorization, we regard the $d\Set F$ as $H_f$ rows of $VL$ elements
and transform the result into original filter size after the final result is acquired.
The examples in Fig. \ref{fig.weight1} and Fig. \ref{fig.weight2} illustrate the computation
procedure of a basic block of our approach.
The relations between rows of $\Set I$, $d\Set O$ and $d\Set F$ are denoted by
the arrows in different colors in the top of the two figures.
In the unity stride case shown in Fig. \ref{fig.weight1},
we first load a row($r_1$) of input and extract it into 2 registers $vi_{0:1}$ by
implicit padding and register manipulations.
Then we multiply $vi_{0:1}$ with the elements of $vo_{0}$ and $vo_{1}$ as indicated by the arrows from $vo_{0:1}$,
and the results are accumulated into $vf_1$ and $vf_0$ respectively.
The results $vf_{0:2}$ will be stored back to $d\Set F_{tmp}$ after the processing of this feature map is completed.
The case of stride=$2$ is shown in Fig. \ref{fig.weight2}.
The loaded row($r_1$) of $\Set I$ is extracted into 2 vectors $vi_{0:1}$ with a stride of 2, as indicated by the numbers.
The multiplications and accumulations are the same as the former case.
We adopt $2 \times 4$ tiling size for the implementation of stride-1 kernels and
$1 \times 4$ tiling size for the implementation of stride-2 kernels.

\subsection{Arithmetic Intensity}
In this part, we will take the forward propagation as an example to
analyze the Arithmetic Intensity (AI) \cite{AI2005} of our optimized implementations and Tengine,
which has the largest AI among the existing implementations.
The total number of arithmetic operations is $TA = 2 \times N \times C \times H_o \times W_o \times H_f \times W_f$.
When only the tiling size $ H_r \times W_r$ is used, the total communication between registers and cache in our approach involves:
\begin{enumerate}
	\item Loading $\Set F$ once for each batch(Line \ref{load_kernel}).
	      So the $\Set F$ incurs $TC_{f}$ = $4 \times N \times C \times H_f \times W_f$ bytes traffic.
	\item Storing $\Set O$ once(Line \ref{store_output}).
	      Thus the traffic of $\Set O$ is $TC_{o}$ = $4 \times N \times C \times H_o \times W_o$ bytes.
	\item Loading $(W_r-1) \times s + W_f$  elements of $\Set I$ in loop r(Line \ref{padding_start}).
	      So the traffic of $\Set I$ incurred by one complete execution of \emph{Kernel}$H_r \times W_r$ is $TC_{ik} = ((W_r-1)\times s+W_f) \times ((H_r-1) \times s + H_f)$.
	      The \emph{Kernel}$H_r \times W_r$ is called $N \times C \times H_o/H_r \times W_o/W_r$ times.
	      Hence the total traffic of $\Set I$ is $TC_{i}$ = $4 \times N \times C \times H_o/H_r \times W_o/W_r \times TC_{ik}$ bytes.
\end{enumerate}
With the tiling size $4 \times 4$, the AI of our implementation is
\begin{equation}
	\begin{aligned}
		AI_{ours} & =\frac{TA}{TC_{f}+TC_{o}+TC{i}} \\
		          & \approx\left\{
		\begin{array}{rc}
			{\frac{1}{( {0.13 + 2/(H_o \times W_o)})}}(ops/Byte) & s=1 \\
			{\frac{1}{( {0.31 + 2/(H_o \times W_o)})}}(ops/Byte) & s=2 \\
		\end{array} \right. 
	\end{aligned}
\end{equation}
The AI of Tengine is:
\begin{equation}
	{\rm{A}}{{\rm{I}}_{tg}} = \frac{TA}{T{C_{tg}}} \approx\left\{
	\begin{array}{rc}
		{\frac{1}{({1.33 + 2/(H_o \times W_o)})}}(ops/Byte) & s=1 \\
		{\frac{1}{({2 + 2/(H_o \times W_o)})}}(ops/Byte)    & s=2 \\
	\end{array} \right. 
\end{equation}
Therefore, the AI of our implementation is larger than that of Tengine.

\section{Experimental Evaluation}
\label{performance}
In this section, we firstly compare our forward propagation implementation to the existing ones in Tengine \cite{tengine2021}, FeatherCNN\cite{feathernn2021}, ncnn\cite{ncnn2021} and ARM Compute Library (ACL) \cite{ArmCL2021}. Secondly, we compare our backward propagation and weight gradient update implementations to the matrix multiplication-based ones in Pytorch. Thirdly, we evaluate the full topology speedup of MobileNetV1 and MobileNetV2 based on our implementations.

\subsection{Experimental Setup}
We run our experiments on the following two ARMv8 processors:

\textbf{Phytium FT1500A\cite{FT1500A2022,chen2018efficient}:} 1.5GHz ARMv8 processor with 2 core groups each with 4 cores. Each core has 32KB L1 instruction cache and 32KB L1 data cache. 4 cores of a core group share 2MB L2 cache.

\textbf{Marvell ThunderX\cite{ThunderXCP}:} 2.0 GHz ARMv8 processor with 48 cores. Each core has 78KB L1 instruction cache and 32KB L1 data cache. All 48 cores share 16MB L2 cache.

In the compilation of Pytorch, we use the OpenBLAS version 0.3.15 library to provide GEMM function.
The depthwise convolutional layers from lightweight networks MobileNetV1 and MobileNetV2 are used in our tests. In the following tables and figures, MobileNetV1 and MobileNetV2 are labeled as v1 and v2.
All the tests are iterated 10 times and the median runtime is reported as the result of each test.
\subsection{Forward Propagation Performance}
We compare our forward propagation implementation against the existing ones in Tengine \cite{tengine2021}, FeatherCNN\cite{feathernn2021}, ncnn\cite{ncnn2021} and ARM Compute Library (ACL) \cite{ArmCL2021} using different cores. Tengine works best in most cases among four open implementations, so we normalize the performance to Tengine.

\begin{figure*}[!htbp]
	\setlength{\abovecaptionskip}{-0.2cm} 
	\setlength{\abovecaptionskip}{0.0cm} 
	\centering
	\subfigure[FT1500A]{
		\centering
		\includegraphics[width=0.45\linewidth]{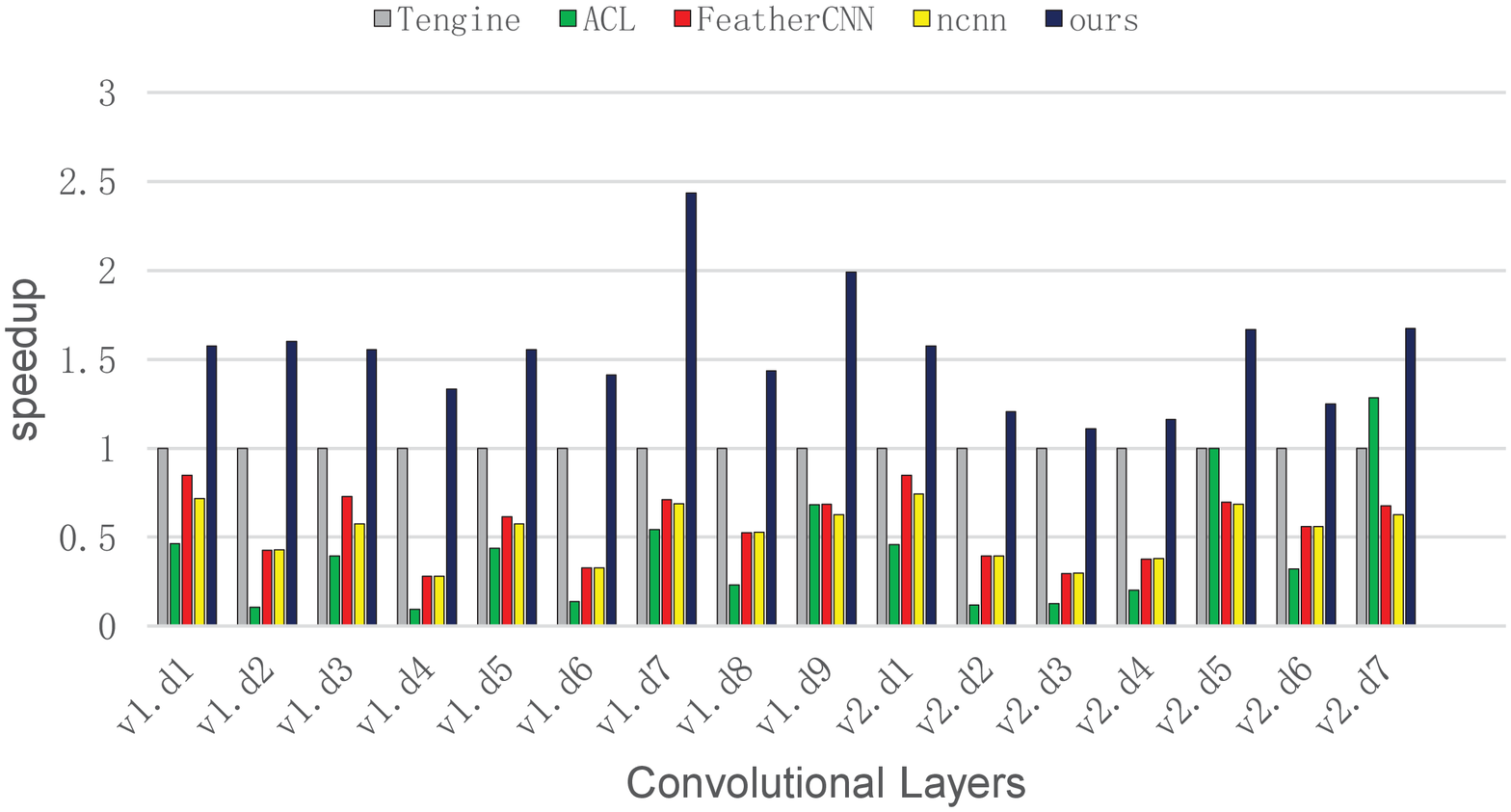}
		\label{ft1500-output-t1b1}}
	\subfigure[ThunderX]{
		\centering
		\includegraphics[width=0.45\linewidth]{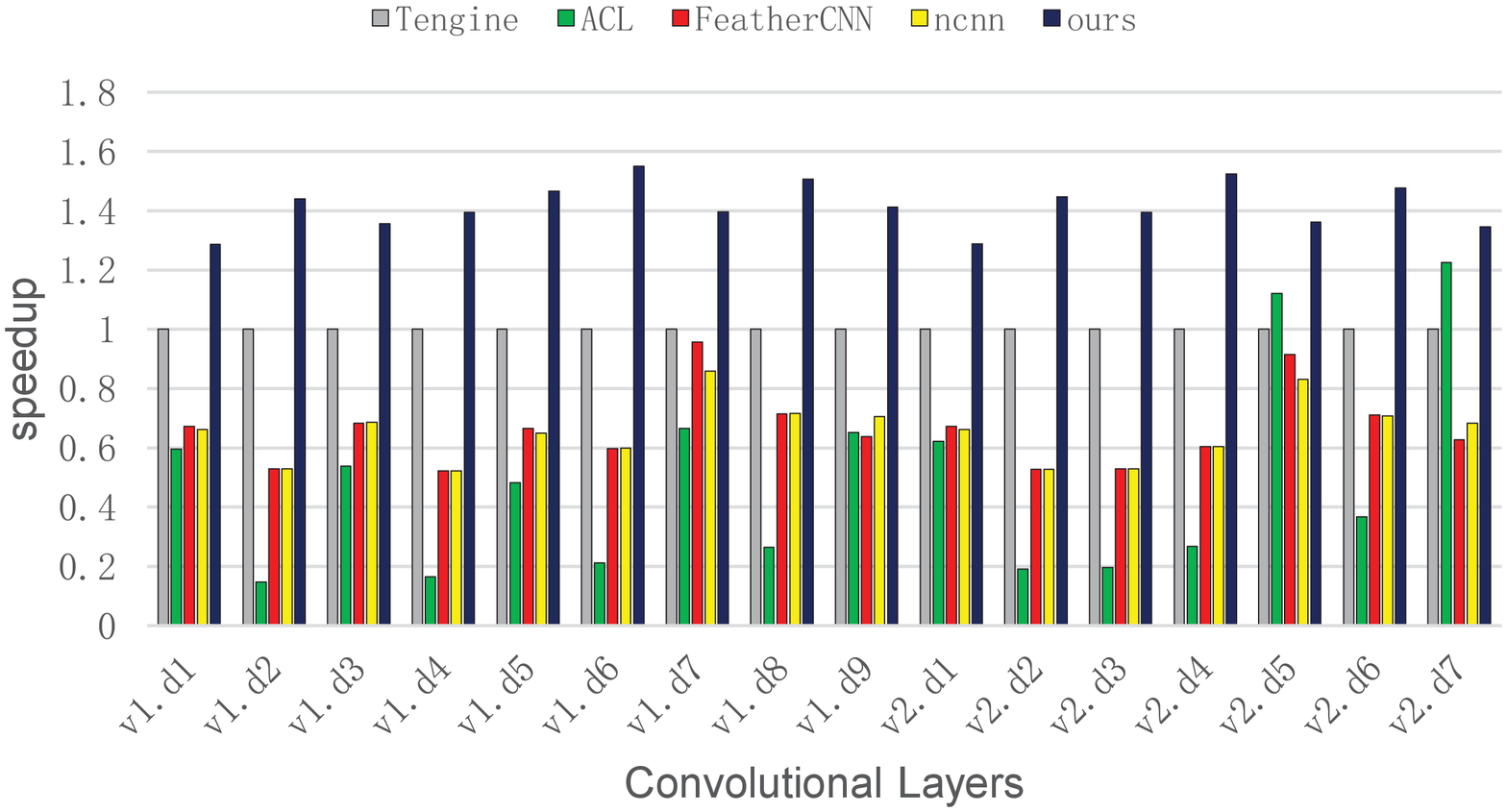}
		\label{thunder-output-t1b1}}
	\caption{Single-core performance of forward propagation implementations on FT1500A and ThunderX, where mini-batch size is 1 and performance is normalized to Tengine.}
\end{figure*}

The relative single-core performance of five implementations on FT1500A and ThunderX is shown in Fig. \ref{ft1500-output-t1b1} and Fig. \ref{thunder-output-t1b1}, respectively. The x-axis indicates the different depthwise convolutional layers from MobileNetV1 and MobileNetV2, while the y-axis shows the speedup of five implementations over Tengine. In the test of the single-core performance, the mini-batch size of all layers is set to 1. The results show that our implementation is better than all four open implementations in all cases on these two ARMv8 platforms. Compared to Tengine, FeatherCNN, ncnn and ACL, our implementation obtains speedups up to 2.43x, 4.75x, 4.30x, 36.38x on FT1500A and 1.54x, 2.72x, 2.74x, 9.80x on ThunderX. When the same input size is adopted, the layers with stride 1 get much bigger speedup than the ones with stride 2. The main reason is that our approach reduces the access of output feature maps as much as possible and the larger output size let our implementation get higher performance improvements.

\begin{figure*}[!htbp]
	\setlength{\abovecaptionskip}{-0.2cm} 
	\setlength{\abovecaptionskip}{0.0cm} 
	\centering
	\subfigure[FT1500A]{
		\centering
		\includegraphics[width=0.45\linewidth]{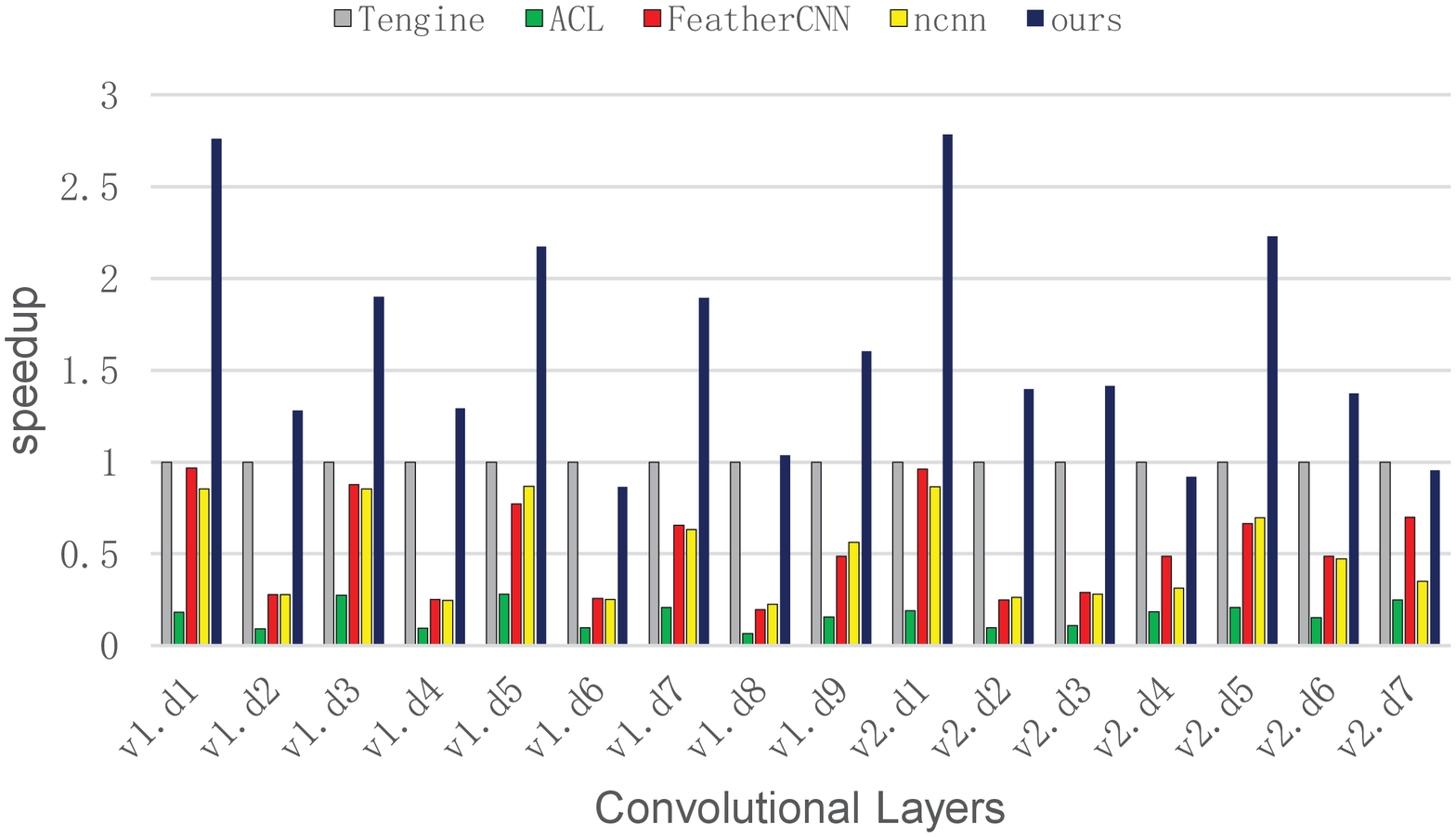}
		\label{ft1500-output-t16b256}}
	\subfigure[ThunderX]{
		\centering
		\includegraphics[width=0.45\linewidth]{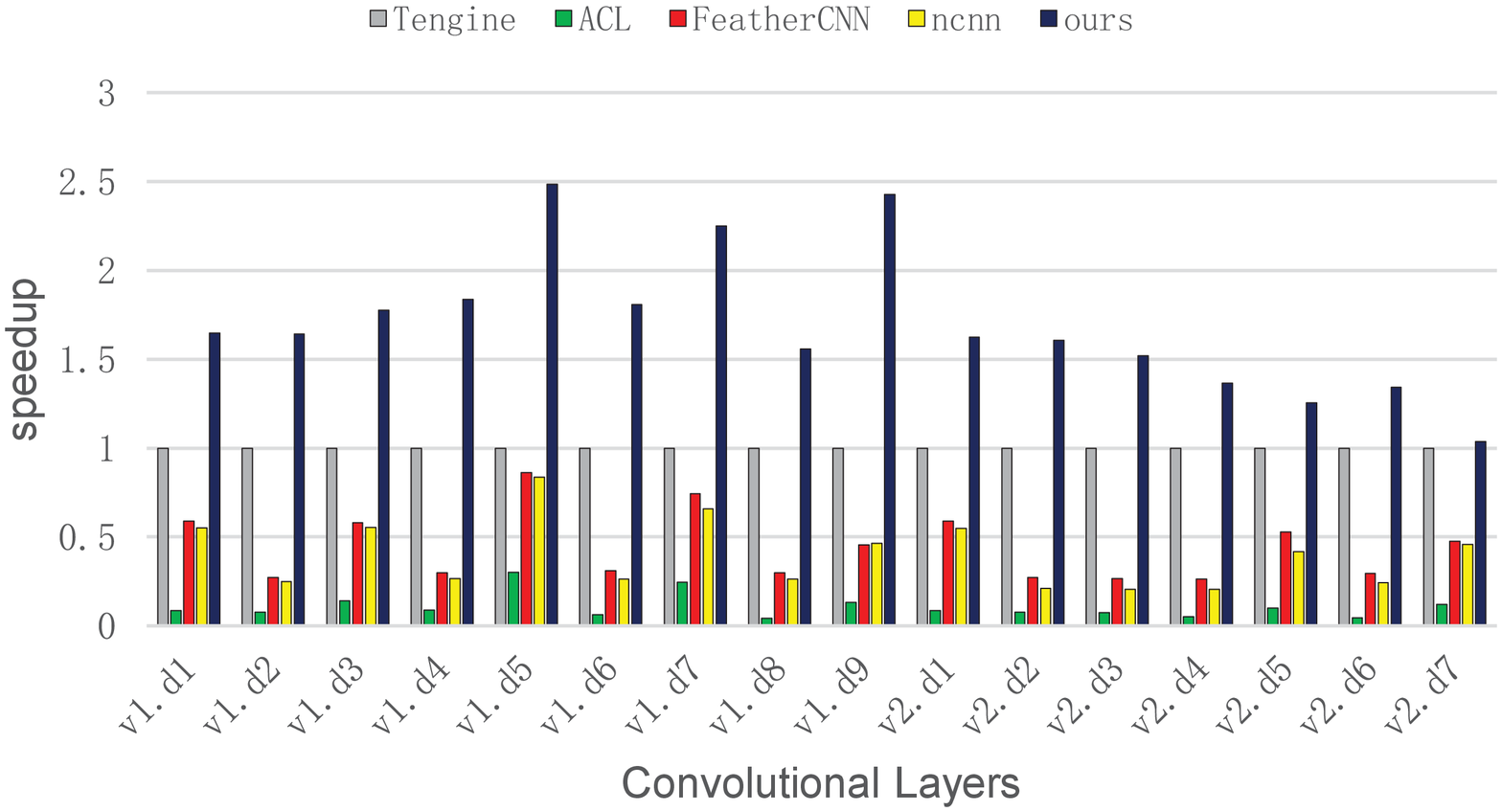}
		\label{thunder-output-t48b256}}
	\caption{Multi-core performance of forward propagation implementations on all 16 cores of FT1500A and 48 cores of ThunderX, where mini-batch size is 256 and performance is normalized to Tengine.}
\end{figure*}

Fig. \ref{ft1500-output-t16b256} and Fig. \ref{thunder-output-t48b256} show the relative multi-core performances of our implementation on FT1500A and ThunderX. We set the mini-batch size to 256 since 256 is frequently used in network training.
The results demonstrate that our approach surpasses FeatherCNN, ncnn, ACL on all tested layers and obtains speedups range from 1.36x to 5.67x, 1.23x to 5.70x, 3.83x to 14.66x on FT1500A and 1.86x to 6.15x, 2.48x to 7.53x, 9.1x to 28.62x on ThunderX. When compared to Tengine, our implementation exhibits higher performance in most tested layers and yields average speedups of 1.55x on FT1500A and 1.57x on ThunderX, respectively.

In summary, our approach can effectively accelerate the forward propagation for depthwise convolutions.


\subsection{Backward Propagation Performance}
\begin{figure*}[!htbp]
	\setlength{\abovecaptionskip}{-0.2cm} 
	\setlength{\abovecaptionskip}{0.0cm} 
	\centering
	\subfigure[FT1500A]{
		\centering
		\includegraphics[width=0.45\linewidth]{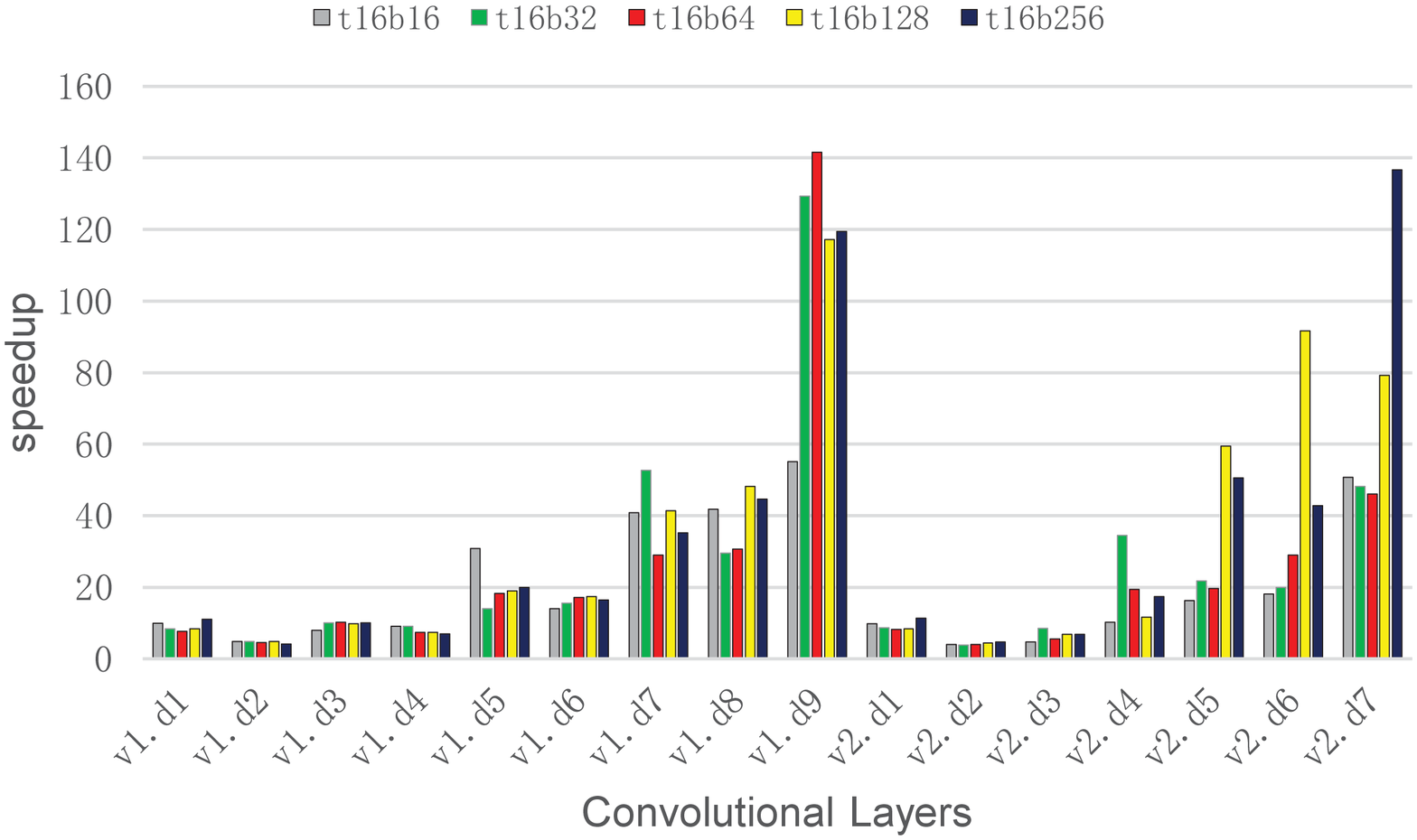} 
		\label{fig-ft1500-input}}
	\subfigure[ThunderX]{
		\centering
		\includegraphics[width=0.45\linewidth]{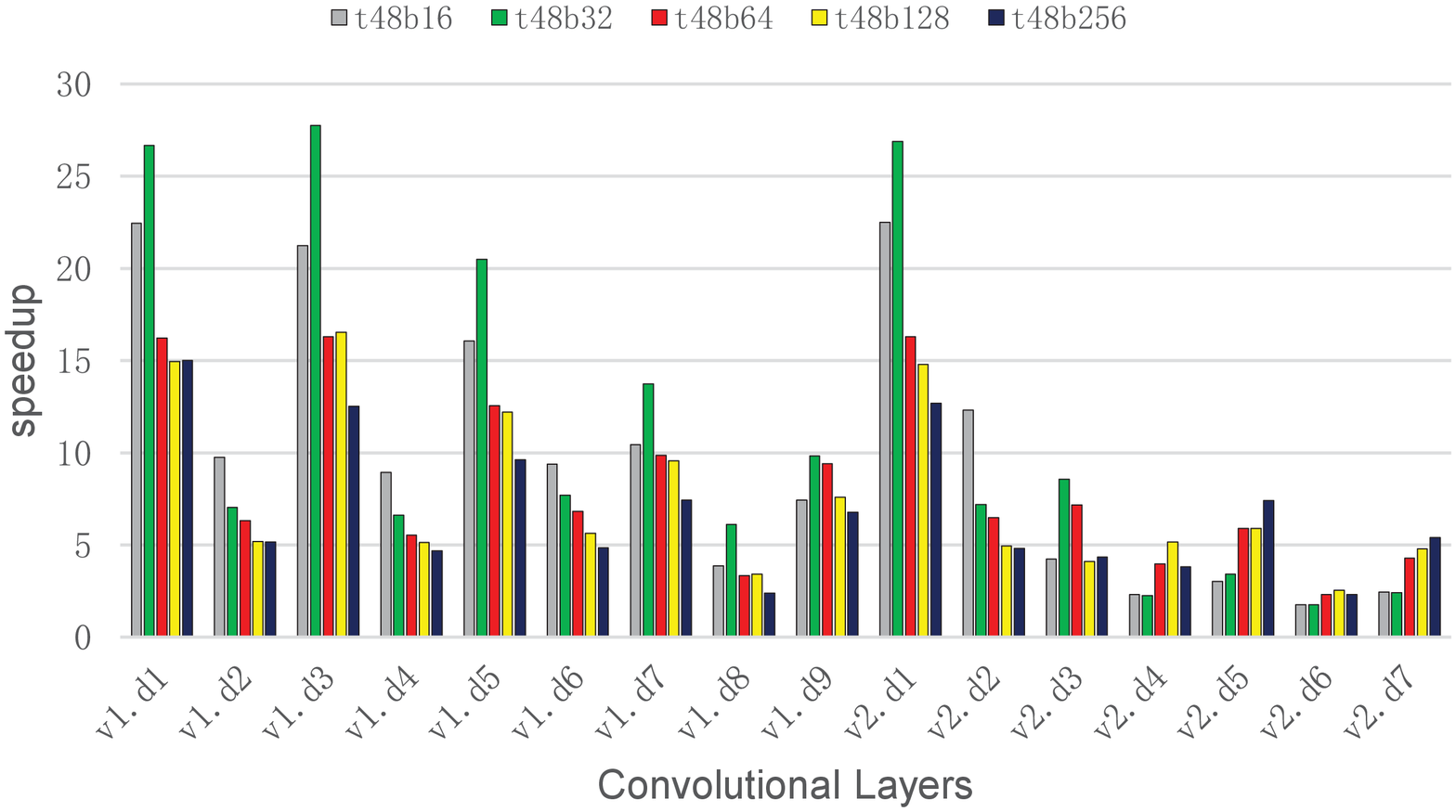} 
		\label{fig-thunder-input}}
	\caption{Speedup of our backward propagation implementation over the one in pytorch on FT1500A and ThunderX, where different mini-batch sizes are used. t?b? represents the thread number and batch size}

\end{figure*}

The relative performance of our backward propagation implementation against the matrix multiplication-based one in pytorch are shown in Fig. \ref{fig-ft1500-input} and Fig. \ref{fig-thunder-input}, respectively. In the figures, different colors represent the tests using different mini-batch sizes. Among all the tests, the speedups of our implementation against Pytorch are greater than 3.96x on FT1500A and 1.76x on ThunderX. For the cases on FT1500A shown in Fig. \ref{fig-ft1500-input}, the speedup increases significantly with the decrease of input size. When input size is small, our tests show that the overhead of matrix multiplication in Pytorch accounts for more than 80\% of the total cost on FT1500A. For the cases on ThunderX shown in Fig. \ref{fig-thunder-input}, the layers with bigger input size show higher speedup. The matrix multiplication's percentage in Pytorch ranges from  31.8\% to 55.7\% on ThunderX. In addition, we can observe that our implementation on FT1500A gets higher average improvement than that on ThunderX. The main reason is that the matrix multiplication-based algorithm is heavily dependent on the performance of functions in BLAS library, and the OpenBLas v0.3.15 library maybe has not been well optimized on FT1500A. In short, our direct implementation for backward propagation of depthwise convolutions are better than the matrix multiplication-based one in Pytorch, and doesn't rely on any external computing libraries.


\subsection{Weight Gradient Update Performance}
\begin{figure*}[!htbp]
	\setlength{\abovecaptionskip}{-2.0cm} 
	\setlength{\abovecaptionskip}{0.0cm} 
	\centering
	\subfigure[FT1500A]{
		\centering
		\includegraphics[width=0.45\linewidth]{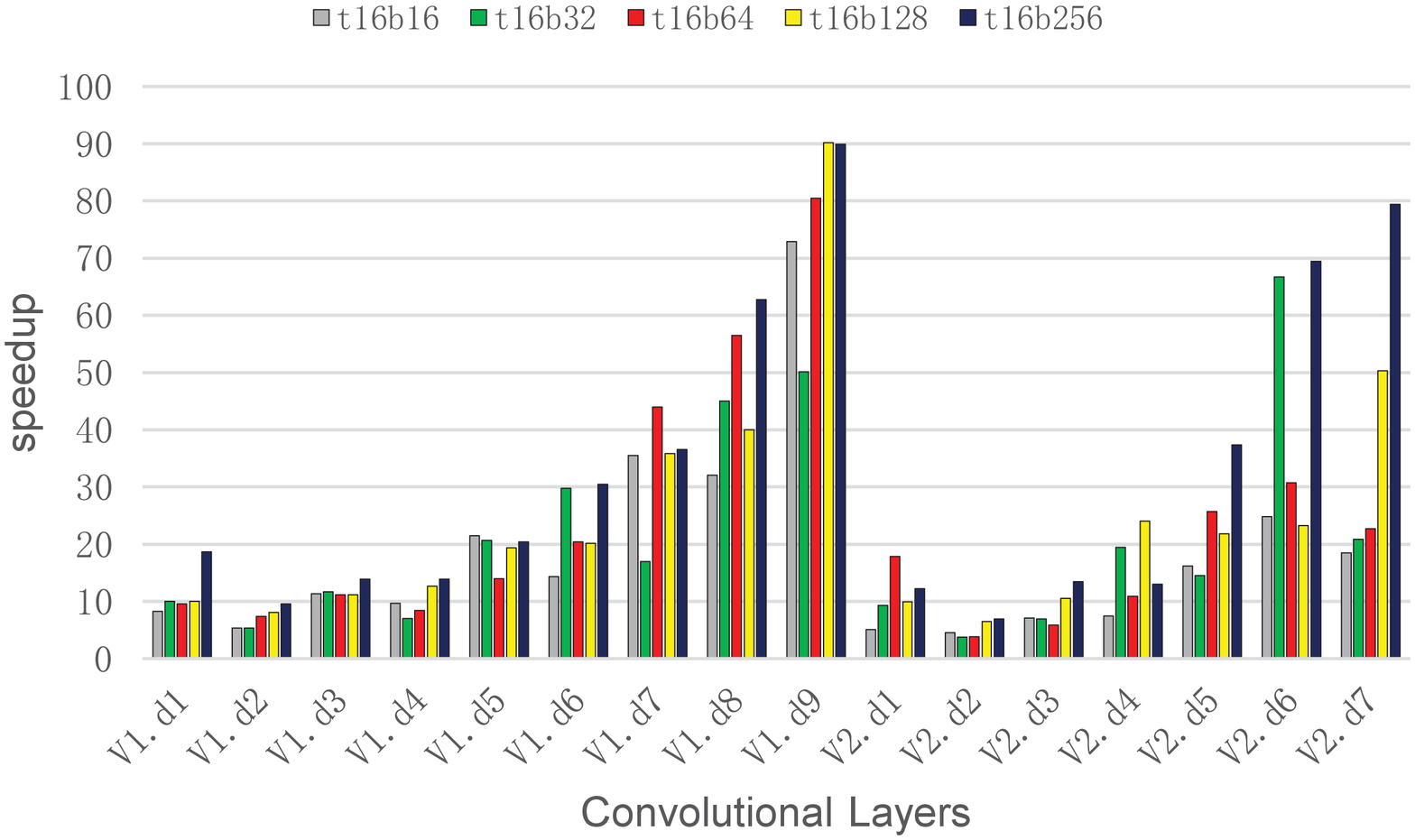} 
		\label{fig-ft1500-kernel}}
	\subfigure[ThunderX]{
		\centering
		\includegraphics[width=0.45\linewidth]{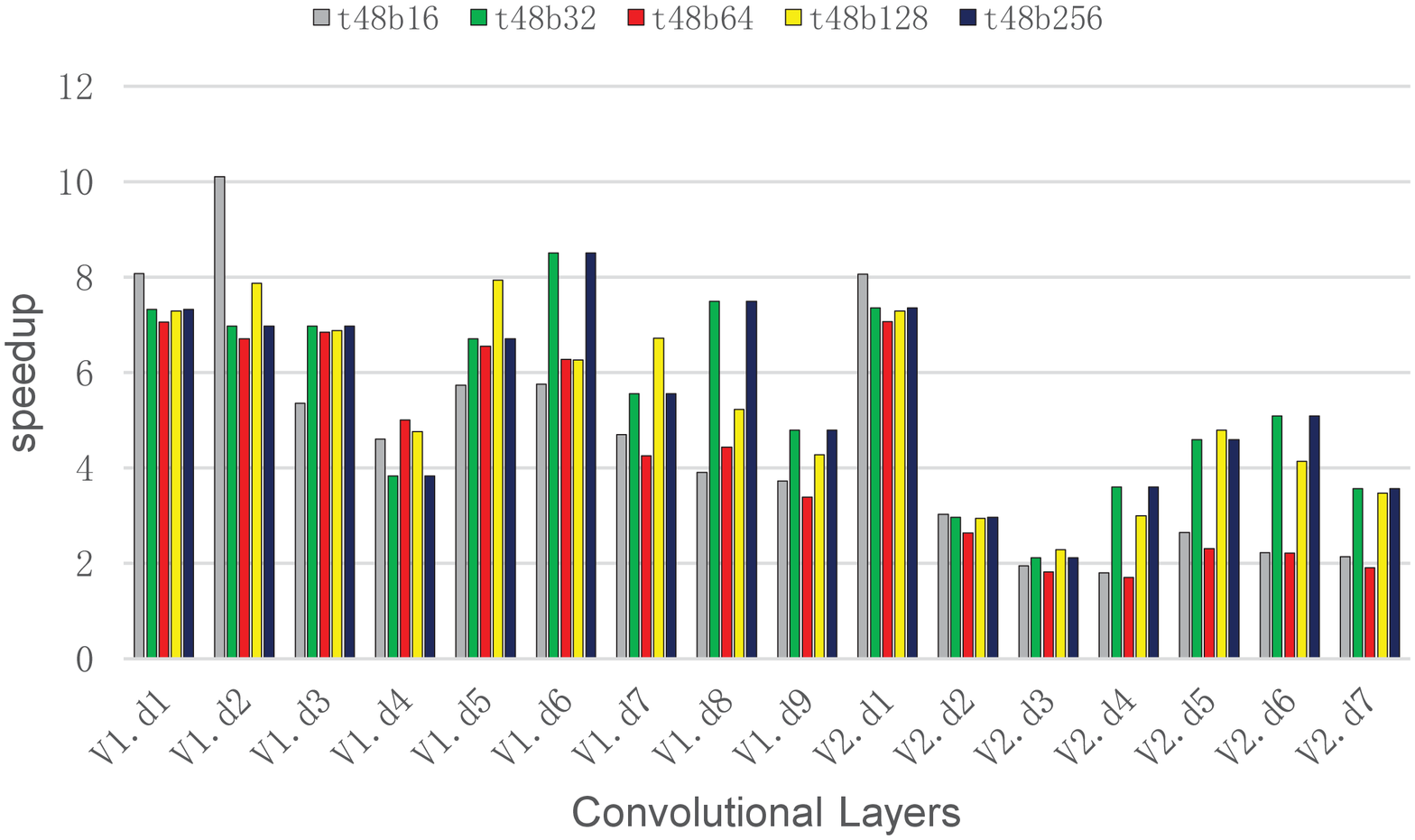} 
		\label{fig-thunder-kernel}}
	\caption{Speedup of our weight gradient update implementation over the one in pytorch on FT1500A and ThunderX, where different mini-batch sizes are used. t?b? represents the thread number and batch size}
\end{figure*}

Fig. \ref{fig-ft1500-kernel} and Fig. \ref{fig-thunder-kernel} show the relative performance of our weight update propagation implementation against the matrix multiplication-based one in pytorch. As described in the previous sub-section, this sub-section carries out the tests using different mini-batch sizes as well. The results show that our implementation acquires speedups of 6.83x - 89.44x on FT1500A and 2.11x - 8.50x on ThunderX against the one in Pytorch. And the performance trend on two CPUs is similar to that for backward propagation.

\subsection{Full Topology Performance}
Finally, we integrated our direct implementations into Pytorch and evaluated the end-to-end inference and training speedup of MobileNetV1 and MobileNetV2 over original Pytorch. The experimental results under different number of threads and mini-batch sizes are provided in Table\ref{table.inference} and Table\ref{table.train} respectively. For the inference of MobileNetV1 and MobileNetV2, our work achieves an average speedup of 4.23x and 3.67x against original Pytorch on FT1500A, and an average speedup of 9.24x and 7.36x against original pytorch on ThunderX. When comparing the training of MobileNetV1 and MobileNetV2 to original Pytorch, an average speedup of 13.72x and 12.96x is observed on FT1500A, and the corresponding one is 23.58x and 21.85x on ThunderX.

\begin{table}[!htbp]
	\footnotesize
	\caption{Inference speedup for MobileNetV1/MobileNetV2 of our work against original pytorch. \textit{t}x denotes that x threads is used and \textit{t}x/y means x threads is used on FT1500A and y threads is used on ThunderX. \textit{b}x indicates that mini-batch size is x.}
	\label{table.inference}
	\begin{tabular}{lcccc}
		\cline{1-5}
		                              & \multicolumn{2}{c}{\textbf{FT1500A}}     & \multicolumn{2}{c}{\textbf{ThunderX}}                                                                                          \\
		                              & \multicolumn{1}{c}{\textbf{MobileNetV1}} & \multicolumn{1}{c}{\textbf{MobileNetV2}} & \multicolumn{1}{c}{\textbf{MobileNetV1}} & \multicolumn{1}{c}{\textbf{MobileNetV2}} \\ \cline{1-5}
		\textit{t}1,\textit{b}1       & 3.04x                                    & 2.62x                                    & 4.26x                                    & 2.38x                                    \\
		\textit{t}16/48,\textit{b}16  & 7.00x                                    & 5.97x                                    & 19.76x                                   & 15.89x                                   \\
		\textit{t}16/48,\textit{b}32  & 5.28x                                    & 4.54x                                    & 13.21x                                   & 10.94x                                   \\
		\textit{t}16/48,\textit{b}64  & 3.74x                                    & 3.66x                                    & 8.39x                                    & 6.90x                                    \\
		\textit{t}16/48,\textit{b}128 & 2.85x                                    & 2.65x                                    & 5.81x                                    & 4.53x                                    \\
		\textit{t}16/48,\textit{b}256 & 3.44x                                    & 2.60x                                    & 4.03x                                    & 3.51x                                    \\
		\cline{1-5}
	\end{tabular}
\end{table}
\begin{table}[!htbp]
	\footnotesize
	\caption{Training speedup for MobileNetV1/MobileNetV2 of our work against pytorch.}
	\label{table.train}
	\begin{tabular}{lcccc}
		\cline{1-5}
		                              & \multicolumn{2}{c}{\textbf{FT1500A}}     & \multicolumn{2}{c}{\textbf{ThunderX}}                                                                                          \\
		                              & \multicolumn{1}{c}{\textbf{MobileNetV1}} & \multicolumn{1}{c}{\textbf{MobileNetV2}} & \multicolumn{1}{c}{\textbf{MobileNetV1}} & \multicolumn{1}{c}{\textbf{MobileNetV2}} \\ \cline{1-5}
		\textit{t}16/48,\textit{b}16  & 15.33x                                   & 15.21x                                   & 18.88x                                   & 19.55x                                   \\
		\textit{t}16/48,\textit{b}32  & 14.03x                                   & 12.73x                                   & 26.89x                                   & 23.26x                                   \\
		\textit{t}16/48,\textit{b}64  & 13.05x                                   & 12.78x                                   & 22.43x                                   & 20.83x                                   \\
		\textit{t}16/48,\textit{b}128 & 12.97x                                   & 12.00x                                   & 25.19x                                   & 21.86x                                   \\
		\textit{t}16/48,\textit{b}256 & 13.20x                                   & 12.09x                                   & 24.53x                                   & 23.73x                                   \\
		\cline{1-5}
	\end{tabular}
\end{table}

\section{Related Work}
\label{relatedwork}
A great amount of work has been made to optimize the implementations of convolution operations. As previously mentioned, the widely-used methods to compute the convolutions contain matrix multiplication-based, Winograd-based, FFT-based and direct algorithms. The first three algorithms can be collectively referred to as the indirect algorithms.

The matrix multiplication-based algorithms, first proposed by Chellapilla \cite{2006High} et al, cast convolutions into matrix-matrix multiplication operations, and can be adopted by convolutions with arbitrary parameters. As a result, the matrix multiplications-based algorithms can be found in all popular deep learning frameworks, such as  Caffe \cite{jia2014caffe}, Mxnet \cite{mxnet2015}, Pytorch \cite{paszke2017pytorch} and TensorFlow \cite{tensorflow2016}. The Winograd- and FFT-based algorithms can reduce the arithmetic complexity of convolutions with specific parameters through Winograd and FFT transformations, and thus they are also called fast convolution algorithms \cite{lavin2016fast, VasilacheJMCPL14}. All the three indirect algorithms have been optimized on ARMv8 architecture \cite{ijcnnwang2019, winograd2020wang, wang2020optimizing, hxd2021optimizing, huangFFT2021}. However, all indirect algorithms generates non-trivial overhead of memory access. Although several methods \cite{ijcnnwang2019, vasudevan2017parallel} have been proposed to optimize the memory access in the matrix multiplications-based algorithms, the overhead is still inevitable. Meanwhile, the fast algorithms maybe increase the computational complexity for depthwise convolutions because their transformations also introduce some computation operations \cite{lavin2016fast}. Thus, all indirect algorithms are not well-suited for depthwise convolutions.

As the direct convolution algorithm often can eliminate all memory overhead, it also attracts a lot of attention \cite{zhang2018high, Evangelos2018, dwaaai2020, dscgpu2022}. Zhang et al. \cite{zhang2018high} and Georganas et al. \cite{Evangelos2018} optimized the direct implementation of conventional convolutions based on specialized layouts, which avoids complex data movement for the vectorization. Lu et al. \cite{dscgpu2022} first proposed an CUDA-based direct implementation of depthwise convolutions on NVIDIA GPUs. Zhang et al. \cite{dwaaai2020} described an optimized forward propagation implementation for depthwise convolutions with NHWC layout on ARM-based mobile devices, which utilized register tiling and loop rescheduling techniques. Unlike \cite{dwaaai2020}, our focus is all the three procedures for depthwise convolutions with NCHW layout on ARMv8 CPUs, which is the default for Caffe, Mxnet, and Pytorch. In theory, depthwise convolutions with NCHW layout can have better data locality than ones with NHWC layout. However, the NCHW layout also increases the memory access overhead accompanied by data alignment in the vectorization on ARMv8 CPUs, which is largely reduced by implicit padding and register tiling in this paper.

\section{Conclusion}
In this paper, we propose new direct implementations of forward propagation, backward propagation, weight gradient update for depthwise convolutions on ARMv8 architectures. Our implementations improve the register-level data locality through implicit padding, vectorization, register tiling and multi-threading techniques so that the communication between cache and registers is optimized. And the arithmetic intensities are analyzed as well. Through the experiments on two ARMv8 CPUs, we show that the new implementations can get better performance than existing implementations and reduce the overhead of end-to-end lightweight CNNs training and inference on ARMv8 CPUs. In the future, we will study how to generate optimized direct implemntations for depthwise convolutions based on just-in-time compilation.

\begin{acks}
	This research work was supported by the National Natural Science Foundation of China under Grant No. 62002365
\end{acks}

\footnotesize
\bibliographystyle{ACM-Reference-Format}
\bibliography{dwconv}
\end{document}